\documentclass[10pt,a4paper]{article}%{revtex4}
\usepackage{hyperref}
\usepackage{amsmath}
\usepackage[psamsfonts]{amssymb }
\usepackage{mathrsfs}

\usepackage{dsfont}

\begin{document}

\title{
Small mass graviton propagator via finite-field-dependent BRST transformations in the critical dimension Siegel-Zwiebach action from string theory
}

\author{Vipul Kumar Pandey${}^1$\footnote {vipulvaranasi@gmail.com} ~and Ronaldo Thibes${}^{2}$\footnote{thibes@uesb.edu.br}\\\\
${}^1$Department of Physics,
Chandigarh University\\
Mohali -- 140413, India\\\\
${}^2$Departamento de Ciências Exatas e Naturais, UESB\\
Rodovia BR 415, km 03, s/n, Itapetinga, Brazil}
\date{November 27, 2023}

\maketitle

\begin{abstract}
We discuss the divergent graviton propagator massless limit problem in $D=26$
and show how it can be rigorously approached by interconnecting distinct gauge-fixed Siegel-Zwiebach generating functionals from string theory in the critical dimension through proper  finite-field-dependent BRST (FFBRST) transformations. The massive Fierz-Pauli Lagrangian can be obtained from the gauge-invariant Siegel-Zwiebach one in the unitary gauge as a particular case, however suffering from the van Dam-Veltman-Zakharov discontinuity and possessing a ill-defined propagator in the massless limit.  Nevertheless, alternatively working in a more suitable generalized Lorenz type gauge, including the transverse-traceless case, the graviton propagator for the Siegel-Zwiebach Lagrangian in the massless limit can be made finite.  Gauge attainability and nilpotent BRST symmetries are explicitly worked out.
We write down the complete corresponding generating functional, including the ghosts sector, and construct a convenient FFBRST transformation connecting the unitary gauge to a new bi-parametrized class of gauge-fixings containing the transverse traceless case.  By taking into account the corresponding change in the Feynman integral Jacobian, a finite massless continuous limit propagator is achieved and fully justified.
\end{abstract}

\section{Introduction} 
After Dirac's initial endeavor towards the quantization of Einstein’s gravity in the canonical Hamiltonian formulation \cite{Dirac:1958sc, PDIR}, followed by Feynman's \cite{RPF} and Weinberg's \cite{SW1,SW2,SW3} attempts,  countless works concerning a quantum formulation for the spin-two particle have been reported in the literature.  In spite of that, a fully satisfactory quantization of gravity still constitutes one of the major open problems in physics.  From the many existing obstacles, we recall Einstein's gravity high nonlinearity along with its nonrenormalization issues and the fact that space-time metric itself becomes a fully dynamical quantum field.
Besides that, from the experimental side, we currently witness a huge growing of unaccountable new facts and data from cosmological observations which reveal an accelerating expanding universe containing unknown forms of dark energy and matter.    In this scenario, clever modifications on General Relativity and on our currently established theoretical frameworks are on demand and certainly deserve the physicist's attention \cite{Reyes:2022dil, Reyes:2022mvm, Pfeifer:2023cgd}.  The particular such possible modification we would like to address here, namely massive gravity, allows for the possibility of an existing non-null small mass for the graviton.  The first and main problem with a massive graviton assumption, however, is the immediate break of gauge symmetry, resulting in a discontinuity of physical predictions as originally pointed out by van Dam, Veltman and  Zakharov \cite{vanDam:1970vg, Zakharov:1970cc} and recently revisited in the context of higher derivatives in \cite{Myung:2017zsa}.  Furthermore, the number of degrees of freedom suffers an abrupt change in the graviton massless limit.  Fair reviews of massive gravity from its inception to current status, including the Vainshtein solution
\cite{Vainshtein:1972sx}
to the van Dam-Veltman-Zakharov discontinuity,  can be seen in \cite{KH,CDR}.   Another well-known proposal in the literature is the ambitious idea of obtaining a consistent description for quantum gravity from string theory with the possible bonus of its unification with other nature fundamental interactions.  
Those two programs can be actually related.  For instance, it is known that higher-spin fields actions can be obtained from string field theory \cite{WSBZ, Bengtsson:1986ys, Sagnotti:2003qa, Asano:2012qn, Asano:2013rka, Lee:2017utr, HPTL}.  It is also worth mentioning that the study of small mass spin-two particles could shed light to a better understanding of the high energy limit for open strings \cite{Gross:1987ar}.

With a credit to string theory, in the current work, we address linearized massive gravity as described by the embedding of the spin-two Fierz-Pauli model \cite{MFWP} into a broader space defined by the gauge-invariant Siegel-Zwiebach (SZ) action \cite{WSBZ} from bosonic string theory in the critical dimension.
In \cite{WSBZ}, Siegel and Zwiebach used a second-quantized BRST-invariant formulation to derive a complete gauge-invariant action describing the bosonic string.  The main functional field lead to auxiliary fields corresponding to higher-spin modes contained in the string and present in Lagrangian functions achieved through expansions 
into
component fields.   More specifically, by expanding the open string action to fourth mass level, Siegel and Zwiebach have obtained a Lagrangian density describing a massive spin-two field interacting with two additional vector and scalar fields \cite{WSBZ}.  That spin-two field, a massive rank two symmetric tensor, can be interpreted as a massive graviton whose properties shall be further explored here.  Due to its origin, in order to enjoy proper gauge invariance, the corresponding massive graviton must abide to the bosonic string critical dimension, living in a $26$-dimensional space-time manifold.
Recently, Park and Lee \cite{HPTL} have proposed the use of the SZ model in a transverse-traceless gauge to study small mass gravitons from string theory.  
Additionally to substantial calculation simplifications, a transverse-traceless gauge has the advantage of propagating only physical degrees of freedom \cite{MTW, Carroll:2004st}.  It has been shown recently that, in the traditional approach to linearized massless gravity, it is not possible to obtain a simultaneously transverse and traceless propagator by means of the usual Faddeev-Popov procedure \cite{BFM}.
Starting from the corresponding classical Lagrangian, 
by using interesting heuristic arguments containing comparisons with the massive Proca model and performing direct substitution of subsidiary gauge conditions back into the SZ Lagrangian itself, Park and Lee have achieved a transverse traceless finite propagator for the massive graviton in the bosonic string context \cite{HPTL}.  Whilst it seems not to be the current case, it is well-known from many examples in quantum field theory that the quantization of gauge-invariant systems can be tricky and sometimes misleading.   Following a different path, our goal here is to extend and justify Park and Lee's idea within a more formal and rigorous functional quantization framework.   More precisely, we apply the BRST quantization technique in its both versions consisting of infinitesimal and finite-field-dependent transformations.  By including the necessary ghost and auxiliary fields, we cast the SZ action into a BRST invariant form assuring unitarity and compute the proper Green functions generating functional within a covariant approach in a broader class of covariant gauges.  Hence, we perform a full-controlled consistent BRST-invariant functional quantization of the SZ model.   Although the gauge-fixed SZ is shown to be left invariant under standard BRST transformations, the functional integration measure variation under finite-field-dependent BRST (FFBRST) transformations has to be taken into account through a corresponding Jacobian \cite{JM1}.  Precisely by means of that nontrivial Jacobian, a FFBRST transformation is shown to connect the unitary and generalized Lorenz type gauges.  In this way, we confirm the consistency of the resulting small mass graviton propagator obtained in the whole class of covariant gauges investigated, including the transverse-traceless case.  As an important extra bonus, we explicitly obtain the expressions for the propagator in a family of generalized Lorenz type gauges depending on two free gauge parameters, which can be indeed very useful in different calculation scenarios.

The present work is organized as follows.  In the next section,  we define our notation and conventions and introduce the Siegel-Zwiebach action describing a massive rank-two tensor interacting with a scalar and a vector fields enjoying gauge invariance in the string critical dimension.   The unitary gauge is found to lead to the massive Fierz-Pauli action as a particular case.  In section {\bf 3}, we work out a two-parameter generalized Lorenz type gauge-fixing for the SZ action in a functional quantization framework.  The transverse-traceless gauge attainability is explicitly shown and can be obtained at quantum level for specific limits of the gauge parameters.  The small mass graviton propagator in the generalized Lorenz gauge is obtained and verified to have a fine well behaved massless transverse-traceless limit.  In section {\bf 4}, we give a brief summary of the FFBRST formalism and proceed to show that the generating functionals in the unitary and generalized Lorenz gauges can be connected by a FFBRST transformation providing a formal proof of their equivalence and assuring the consistency of the previous results.  We end in section {\bf 5} with some concluding remarks.

\section{Gauge-invariant massive Siegel-Zwiebach action}
In string theory, massive spin-two fields arise in the spectrum of open strings as part of a corresponding rank-two tensor multiplet.
In fact, as mentioned in the Introduction, starting from a BRST-invariant formulation for the free bosonic string and expanding its functional in terms of component fields, Siegel and Zwiebach  have obtained the $D=26$ space-time dimension gauge invariant massive spin-two action
\begin{eqnarray}
{S}_{SZ} &=& \int d^{26}x\, \frac{1}{2}\left\{ \frac{1}{2}h_{\mu\nu}{(\Box - m^2)}h^{\mu\nu} + B_\mu{(\Box - m^2)} B^\mu - \theta{(\Box - m^2)} \theta \right. \nonumber \\&& \left. + {\left(\partial^\nu h_{\mu\nu} + \partial_\mu \theta - m B_\mu\right)^2} + {\left(\frac{mh}{4} + \frac{3m\theta}{2} + \partial_\mu B^\mu\right)^2} \right\}
\label{LSZ}
\end{eqnarray}
describing a mass $m$ symmetric tensor field $h_{\mu\nu}$, with trace denoted by $h \equiv h^{\mu}_{\;\mu}$, coupled to two auxiliary vector and scalar fields, $B^\mu$ and $\theta$  \cite{WSBZ}.
Given 27 arbitrary space-time dependent parameters $\epsilon$ and $\epsilon^\mu$,
the above SZ action (\ref{LSZ}) can be readily verified to be invariant under the local gauge transformations\footnote{Constant flat Minkowsky space-time metric is denoted by $\eta^{\mu\nu}$ with signature convention diag $\eta^{\mu\nu}=(-1,1,\dots,1)$ and Greek indexes running through $0,1,\dots,25$.}
\begin{equation}
\begin{split}
\delta h_{\mu\nu} &= \partial_\mu \epsilon_{\nu} + \partial_\nu \epsilon_{\mu} - \frac{1}{2}m\eta_{\mu\nu}\epsilon\,,\\
\delta B_{\mu} &= \partial_\mu \epsilon + m \epsilon_\mu\,,\\
\delta \theta &= - \partial_\mu \epsilon^\mu + \frac{3}{2}m\epsilon\,,
\end{split}
\label{gs}
\end{equation}
holding exactly for the critical dimension $D=26$.
This can be understood by tracing back the gauge symmetry (\ref{gs}) to its  origin corresponding to the nilpotency of the BRST charge in open bosonic string theory, which is also required for the absence of anomalies at quantum level.  
The canonical constraint analysis of (\ref{LSZ}) has been recently studied in reference \cite{HPTL}. 
Note that, if we substitute 
\begin{eqnarray}
B^\mu = 0, \quad \theta = -\frac{h}{2}\,,
\label{FPG}
\end{eqnarray}
directly into (\ref{LSZ}) we obtain
\begin{equation}
{S}_{FP} = \int d^{26}x\, \left\{ \frac{1}{4}h_{\mu\nu}{(\Box - m^2)}h^{\mu\nu}
+ \frac{1}{2} \partial^\nu h_{\mu\nu} \left( \partial^\rho h^\mu_{\;\rho}-\partial^\mu h \right)
  - \frac{1}{4} h{(\Box - m^2)} h \right\} \,,
\label{SFP}
\end{equation}
which happens to be the non-gauge-invariant massive Fierz-Pauli action \cite{MFWP}.  In this way, $B^\mu$ and $\theta$ can be interpreted as 
Stueckelberg fields \cite{Stueckelberg:1957zz} reassuring gauge-invariance for the massive case.  Furthermore, for $m=0$ in (\ref{SFP}), we recover the linearized Einstein-Hilbert action with its usual invariance under
\begin{equation}
\delta h_{\mu\nu} = \partial_\mu \epsilon_{\nu} + \partial_\nu \epsilon_{\mu} 
\,.
\end{equation} 
This suggests that (\ref{FPG}) should be a suitable subsidiary gauge-fixing condition for (\ref{LSZ}), playing the role of a unitary gauge relating two corresponding first- and second-class systems \cite{Amorim:1999xr}.  In fact, given an arbitrary field configuration $(h^{\mu\nu}, B^\mu, \theta)$, by choosing the gauge parameters
\begin{equation}
\begin{split}
\epsilon &=
\frac{1}{5m}\left(\theta+\frac{h}{2}\right)\,,
\\
\epsilon^\mu &=-\frac{B^\mu}{m}
- \frac{1}{5m^2}\partial^\mu\left(\theta+\frac{h}{2}\right)\,,
\end{split}
\end{equation}
it is always possible to achieve (\ref{FPG}).  However, in order to rigorously implement this result at quantum level, one needs to write down the complete Green's functions generating functional, particularly paying attention to integration measure effects coming from constraints and gauge freedom.  In a BRST quantization framework  \cite{Becchi:1974xu, Becchi:1974md, Tyutin:1975qk}, that can be done by first introducing the anticommuting ghost-antighost fields, $(c, \bar{c})$ and $(c^\mu,\bar{c}_\mu)$, the Nakanish-Lautrup auxiliary fields $(b,b^\mu)$, defining the full nilpotent BRST transformations from
\begin{equation}
\begin{array}{c}
s h_{\mu\nu} = \partial_\mu c_{\nu} + \partial_\nu c_{\mu} - \displaystyle\frac{m}{2}\eta_{\mu\nu}c\,,\\[10pt]
s B_{\mu} = \partial_\mu c + m c_\mu\,,\quad
s \theta = - \partial_\mu c^\mu + \displaystyle\frac{3}{2}m c \,,\\[10pt]
s c_{\mu} = 0 \,,\quad
s \bar{c}_{\mu} = b_\mu\,,\quad
\quad s b_\mu = 0\,,
 \\[10pt] 
s c = 0 \,,\quad 
s \bar{c} = b \,,\quad 
s b = 0\,, 
\end{array}
\label{BRST}
\end{equation}
and then adding the gauge-fixing and ghosts terms, conveniently written as
\begin{equation}
S_{ug} = \int d^{26}x\, 
\left[b_\mu B^\mu + b\theta +\frac{bh}{2}
-{5m}\bar{c}c+\bar{c}_\mu \partial^\mu c + m \bar{c}_\mu c^\mu 
\right]
\,,
\end{equation}
to the classical action (\ref{LSZ}).  Now it can be checked that the generating functional
\begin{equation}
Z_u = \int d\mu \, e^{i \displaystyle \left\{ {S}_{SZ} + S_{ug} \right\}}
\label{Z0}
\end{equation}
with integration measure
\begin{equation}
d\mu \equiv [dh_{\mu\nu}][dB_\mu][d\theta][dc][dc_\mu]
[d\bar{c}][d\bar{c}_\mu][db][db_\mu]
\label{IM}
\end{equation}
is left invariant under BRST transformations constructed from (\ref{BRST}) multiplied by a global infinitesimal anticommuting parameter.  From (\ref{Z0}), after functional integrations in $b$, $\theta$, $b^\mu$ and $B^\mu$, we may calculate the massive spin two field propagator in the unitary gauge as
\begin{eqnarray}
G^{\alpha\beta \sigma\lambda} &=& \frac{i}{p^2 + m^2}\Bigg\{ \frac{2}{25}\left( \eta^{\alpha\beta} + \frac{p^\alpha p^\beta}{m^2}\right)\left( \eta^{\sigma\lambda} + \frac{p^\sigma p^\lambda}{m^2}\right) \nonumber\\&& - \left( \eta^{\alpha\sigma} + \frac{p^\alpha p^\sigma}{m^2}\right)\left( \eta^{\beta\lambda} + \frac{p^\beta p^\lambda}{m^2}\right) - \left( \eta^{\alpha\lambda} + \frac{p^\alpha p^\lambda}{m^2}\right)\left( \eta^{\beta\sigma} + \frac{p^\beta p^\sigma}{m^2}\right)   \Bigg \}\,,
\label{PPFPL}
\end{eqnarray}
satisfying
\begin{equation}
G^{\alpha\beta \sigma\lambda} = G^{\beta\alpha \sigma\lambda} = G^{\alpha\beta \lambda\sigma}
= G^{\sigma\lambda \alpha\beta}
\,,
\end{equation}
describing the massive graviton.  Thus we have checked the consistency of the gauge-fixing (\ref{FPG}) applied to (\ref{LSZ}) at quantum level, as it leads to the proper massive Fierz-Pauli propagator.  However, as can be immediately seen from (\ref{PPFPL}), the sensible massless limit is not well-defined.   Since we are interested in small mass gravitons, we may take advantage that we have a more general action (\ref{LSZ}) and use its allowed gauge freedom to work with an alternative more suitable gauge-fixing.  That is the main task of next section.

\section{Towards a Functional Transverse-Traceless Gauge-Fixing}
We have seen that, whilst consistent, the unitary gauge leads to a ill-defined massless propagator for the graviton.  On the other hand, we next show that a generalization of the usual Lorenz condition in QED for the spin-two field  $h_{\mu\nu}$, along with a trace nullity demand, respectively expressed as
\begin{equation}
\partial_\mu h^\mu_{\,\,\nu} = 0 \mbox{~~~and~~~} h=0 \,,
\label{TTgauge}
\end{equation}
does a better job.  More precisely, conditions (\ref{TTgauge}) can be understood as a transverse-traceless gauge-fixing \cite{MTW}.  The attainability of (\ref{TTgauge}) in the current Siegel-Zwiebach model can be proven if we manage to find gauge transformation parameters simultaneously satisfying
\begin{equation}
\partial_\mu \epsilon^\mu -\frac{13m}{2}\epsilon = -h/2
\label{e1}
\end{equation}
and
\begin{equation}
\Box \epsilon_\mu + \partial_\mu \partial_\nu \epsilon^\nu
-\frac{m}{2}\partial_\mu \epsilon = - \partial_\nu h^\nu_{\,\mu}
\,.
\label{e2}
\end{equation}
In fact,
given an arbitrary field configuration, by choosing
\begin{equation}
\epsilon = \frac{2}{25m}
\left[
h-\Box^{-1}\partial^\mu\partial^\nu h_{\mu\nu}
\right]
\end{equation}
and
\begin{equation}
\epsilon^\mu = \frac{12}{25}
\Box^{-2}\partial^\mu\partial^\rho\partial^\nu h_{\rho\nu}
+\frac{1}{50}\Box^{-1}\partial^\mu h
-\Box^{-1}\partial_\nu h^{\mu\nu}
\,,
\end{equation}
which represents a solution to equations (\ref{e1}) and (\ref{e2}), 
it is always possible to perform a gauge transformation
\begin{equation}
h_{\mu\nu} \rightarrow h'_{\mu\nu}= h_{\mu\nu} +\partial_\mu \epsilon_{\nu} + \partial_\nu \epsilon_{\mu} - \frac{1}{2}m\eta_{\mu\nu}\epsilon\,,
\end{equation}
such that 
\begin{eqnarray}
\partial_\mu h'^\mu_{\,\,\,\nu} = 0, \quad h' = 0
\,,
\end{eqnarray}
establishing the attainability of (\ref{TTgauge}).
At quantum level,
similarly to the Landau gauge in QCD and its Abelian version in QED, we may realize (\ref{TTgauge}) as convenient limits of $R_\xi$ and $R_\zeta$ linear covariant  gauges in a functional quantization framework in terms of two gauge fixing parameters $\xi$ and $\zeta$, fully exploring the BRST symmetry generated from (\ref{BRST}), along the lines of reference \cite{Mandal:2022zil}.  In this way, we write the fundamental generating functional
\begin{equation}
Z = \int d\mu \, e^{i \displaystyle  S_{tt} }
\label{Z1}
\end{equation}
with a corresponding gauge-fixed quantum action
\begin{equation}\label{Stt}
S_{tt} = {S}_{SZ} + S_{tg}
\,,
\end{equation}
given by the sum of $S_{SZ}$ as in equation (\ref{LSZ}) and
\begin{eqnarray}
S_{tg} &=&
\int d^{26}x \left\{
\frac{\zeta b^2}{2m^2} + bh +\frac{\xi}{2}b^\mu b_\mu + b^\mu \partial_\nu h^\nu_{\,\mu}
+{\bar{c}}^\mu\Box c_\mu + \bar{c}^\mu \partial_\mu \partial_\nu c^\nu 
\right.\nonumber\\&&\left.
-\frac{m}{2} \bar{c}^\mu \partial_\mu c
+2 \bar{c} \partial_\mu c^\mu - 13 m \bar{c}c
\right\}
\,.
\end{eqnarray}
The integration measure $d\mu$ in (\ref{Z1}) is the same as given by (\ref{IM}).  The independence of the gauge implementation in the generating functional within the BRST framework, so that (\ref{Z0}) and (\ref{Z1}) describe the same physics, will be proven in the next section in terms of FFBRST transformations. 
The quantum action (\ref{Stt}) can be verified to be invariant under BRST transformations
\begin{equation}\label{BRS}
\phi \longrightarrow \phi+(s\phi)\omega
\end{equation}
 generated 
by equations (\ref{BRST})
with
\begin{equation}
\phi\equiv(h_{\mu\nu},B_\mu,\theta,b,b_\mu,c,c_\mu,\bar{c},\bar{c}_\mu)
\end{equation}
denoting a generic field and $\omega$ a infinitesimal anticommuting global parameter.
By performing a functional integration in (\ref{Z1}) over the field variables $b$, $b^\mu$ and $\theta$, we obtain the partial result
\begin{equation}
Z = \int d\mu' \, e^{i \displaystyle  S' }
\label{Z2}
\end{equation}
with
\begin{equation}
d\mu' \equiv [dh_{\mu\nu}][dB_\mu][dc][dc_\mu]
[d\bar{c}][d\bar{c}_\mu]
\,,
\end{equation}
\begin{eqnarray}
S'&=&
\int d^{26}x 
\left\{
\frac{1}{4}h_{\mu\nu}(\Box - m^2)h^{\mu\nu} + \frac{1}{2}B_\mu(\Box - m^2)B^\mu + \frac{1}{2}\left( \partial^\nu h_{\mu\nu} - m B_\mu \right)^2
+\frac{1}{2}\left(\frac{mh}{4}+\partial_\mu B^\mu\right)^2 
\right.\nonumber\\&&\left.
-\frac{1}{2}
\left[ \frac{5m}{2}\partial_\mu B^\mu - \partial_\mu \partial_\nu h^{\mu\nu} +\frac{3m^2h}{8}\right]
{\cal M}^{-1} \left[ \frac{5m}{2}\partial_\mu B^\mu - \partial_\mu \partial_\nu h^{\mu\nu} +\frac{3m^2h}{8}\right]
-\frac{m^2h^2}{2\zeta}
\right.\nonumber\\&&\left.
-\frac{1}{2\xi}(\partial^\nu h_{\mu\nu})^2
+{\bar{c}}^\mu\Box c_\mu + \bar{c}^\mu \partial_\mu \partial_\nu c^\nu 
-\frac{m}{2} \bar{c}^\mu \partial_\mu c
+2 \bar{c} \partial_\mu c^\mu - 13 m \bar{c}c
\right\}
\,,
\end{eqnarray}
and
\begin{equation}
{\cal M}\equiv -2 \Box +\frac{13m^2}{4}
\,.
\end{equation}
Further functional integration over $B_\mu$ in equation (\ref{Z2}) leads to
\begin{equation}
Z = \int [dh_{\mu\nu}][dc][dc_\mu]
[d\bar{c}][d\bar{c}_\mu] \, e^{i \displaystyle  S_{eff} }
\label{Zeff}
\end{equation}
in terms of a quantum effective action
\begin{eqnarray}\label{Seff}
S_{eff}
&=& 
\int d^{26}x 
\left\{
\frac{1}{4}h_{\mu\nu}(\Box-m^2)h^{\mu\nu} +\frac{1}{2}(\partial^\nu h_{\mu\nu})^2 +\frac{m^2h^2}{32}
-\frac{m^2}{2} T^\mu {\cal N}^{}_{\mu\nu} T^\nu 
\right.\nonumber\\&&\left.
-\frac{1}{2}
\left[ \partial_\mu \partial_\nu h^{\mu\nu} - \frac{3m^2h}{8} \right]
{\cal M}^{-1}
\left[ \partial_\rho \partial_\lambda h^{\rho\lambda} - \frac{3m^2h}{8} \right]
-\frac{m^2h^2}{2\zeta}
\right.\nonumber\\&&\left.
-\frac{1}{2\xi}(\partial^\nu h_{\mu\nu})^2
+{\bar{c}}^\mu\Box c_\mu + \bar{c}^\mu \partial_\mu \partial_\nu c^\nu 
-\frac{m}{2} \bar{c}^\mu \partial_\mu c
+2 \bar{c} \partial_\mu c^\mu - 13 m \bar{c}c
\right\}
\,,
\end{eqnarray}
where
\begin{equation}
T^\mu \equiv
\partial_\nu h^{\mu\nu} +\frac{\partial^\mu h}{4}
-\frac{5}{2}{\cal M}^{-1}\partial^\mu
\left(\frac{3m^2h}{8}-\partial_\rho\partial_\nu h^{\rho\nu}\right)
\,,
\end{equation}
and
\begin{equation}
{\cal N}^{\mu\nu}\equiv
\Box^{-1} \eta^{\mu\nu}
+\left(\frac{4}{25m^2}{\cal M}^{}-\mathds{1}\right) \Box^{-2}\partial^\mu \partial^\nu
\,.
\end{equation}
After integrations by parts, the effective action (\ref{Seff}) can be rewritten as
\begin{equation}
S_{eff} = 
\int d^{26}x\,
\left\{
\frac{1}{2}
h_{\mu\nu}
{\cal O}^{\mu\nu\alpha\beta}
h_{\alpha\beta}
+{\bar{c}}^\mu\Box c_\mu + \bar{c}^\mu \partial_\mu \partial_\nu c^\nu 
-\frac{m}{2} \bar{c}^\mu \partial_\mu c
+2 \bar{c} \partial_\mu c^\mu - 13 m \bar{c}c
\right\}
\end{equation}
with
\begin{eqnarray}
{\cal O}^{\mu\nu\alpha\beta} &\equiv&
\frac{\Box-m^2}{4}
\left( \eta^{\mu\alpha}\eta^{\nu\beta}+\eta^{\mu\beta}\eta^{\nu\alpha} \right)
+\frac{12\Box^{-2}}{25}\left(\Box-m^2\right)
\partial^\mu \partial^\nu \partial^\alpha \partial^\beta
\nonumber\\&&
+\frac{1}{4}\left[
\frac{1}{\xi}-\Box^{-1}(\Box-m^2)
\right]
\left(
  \eta^{\mu\alpha}\partial^\nu\partial^\beta
+\eta^{\nu\alpha}\partial^\mu\partial^\beta
+\eta^{\mu\beta}\partial^\nu\partial^\alpha
+\eta^{\nu\beta}\partial^\mu\partial^\alpha
\right)
\nonumber\\&&
-\left[
\frac{\Box-m^2}{25}+\frac{m^2}{\zeta}
\right]
\eta^{\mu\nu}\eta^{\alpha\beta}
+\Box^{-1}\left[  \frac{\Box-m^2}{50}  \right]
\left(
\eta^{\mu\nu}\partial^\alpha \partial^\beta+
\eta^{\alpha\beta}\partial^\mu \partial^\nu
\right)
\end{eqnarray}
from which we may compute the massive graviton propagator in momentum space as
\begin{eqnarray}\label{G}
G_{\mu\nu\alpha\beta}^{\xi\zeta}&=&
-\frac{i\xi}{p^2}
\left[
\frac{p_\mu p_\nu p_\alpha p_\beta}{p^4}
+4\left(\eta_{\mu\alpha}-\frac{p_\mu p_\alpha}{p^2}\right)\frac{p_\beta p_\nu}{p^2}
-\frac{2}{25}\left(\eta_{\mu\nu}-\frac{p_\mu p_\nu}{p^2}\right)\frac{p_\alpha p_\beta}{p^2}
\right.\nonumber\\&&\left.
+F^{\xi\zeta}(p^2,m^2)\left(\eta_{\mu\nu}-\frac{p_\mu p_\nu}{p^2}\right)\left(\eta_{\alpha\beta}-\frac{p_\alpha p_\beta}{p^2}\right)
\right]
\nonumber\\&&
-\frac{2}{p^2+m^2}
\left[
\left(\eta_{\mu\alpha}-\frac{p_\mu p_\alpha}{p^2}\right)\left(\eta_{\nu\beta}-\frac{p_\nu p_\beta}{p^2}\right)
-\frac{1}{25}\left(\eta_{\mu\nu}-\frac{p_\mu p_\nu}{p^2}\right)\left(\eta_{\alpha\beta}-\frac{p_\alpha p_\beta}{p^2}\right)
\right]
\,,
\end{eqnarray}
with
\begin{equation}
F^{\xi\zeta}(p^2,m^2)\equiv\frac{\xi}{1250p^2} \left[
\frac{p^2+m^2-50m^2\zeta^{-1}-50p^2\xi^{-1}}{p^2+m^2-50m^2\zeta^{-1}}
\right]
\,.
\end{equation}
To save space and enhance clarity, equation  (\ref{G}) is written in a compact form
where it is understood that all Lorentz indexes should be properly symmetrized under $\mu\leftrightarrow\nu$, $\alpha\leftrightarrow\beta$ and $(\mu\nu)\leftrightarrow(\alpha\beta)$.  Therefore, $G_{\mu\nu\alpha\beta}^{\xi\zeta}$ enjoys the usual symmetry properties
\begin{equation}
G^{\xi\zeta}_{\mu\nu \alpha\beta} = G^{\xi\zeta}_{\nu\mu \alpha\beta} = G^{\xi\zeta}_{\mu\nu \beta\alpha}
= G^{\xi\zeta}_{\alpha\beta \mu\nu}
\,.
\end{equation}
With different choices for the parameters pair $(\xi,\zeta)$, we may realize different specific gauges suitable for concrete calculations.  In fact, equation  (\ref{G}) furnishes the propagator for an extensive class of $(\xi,\zeta)$-covariant gauges.   In particular, in the limits $\xi\rightarrow0$, $\zeta\rightarrow0$ we obtain the fine propagator
\begin{eqnarray}\label{G00}
G_{\mu\nu\alpha\beta}^{00}
&=&
-\frac{i}{p^2+m^2}
\left[
\left(\eta_{\mu\alpha}-\frac{p_\mu p_\alpha}{p^2}\right)\left(\eta_{\nu\beta}-\frac{p_\nu p_\beta}{p^2}\right)
+
\left(\eta_{\mu\beta}-\frac{p_\mu p_\beta}{p^2}\right)\left(\eta_{\nu\alpha}-\frac{p_\nu p_\alpha}{p^2}\right)
\right.\nonumber\\&&\left.
-\frac{2}{25}\left(\eta_{\mu\nu}-\frac{p_\mu p_\nu}{p^2}\right)\left(\eta_{\alpha\beta}-\frac{p_\alpha p_\beta}{p^2}\right)
\right]
\end{eqnarray}
which is perfectly well defined in the massless limit and satisfies the traceless-transverse properties
\begin{equation}
p^\mu G_{\mu\nu\alpha\beta}^{00} = p^\nu G_{\mu\nu\alpha\beta}^{00} = p^\alpha G_{\mu\nu\alpha\beta}^{00} = p^\beta G_{\mu\nu\alpha\beta}^{00} = 0 
\end{equation}
and
\begin{equation}
 \eta^{\mu\nu}G_{\mu\nu\alpha\beta}^{00} = \eta^{\alpha\beta}G_{\mu\nu\alpha\beta}^{00} = 0
\,.
\end{equation}
However,
in order to attest the consistency of results (\ref{G}) and (\ref{G00}), we must investigate the relation between the quantum theories resulting from the previous actions.  Since they correspond to distinct gauge choices, we may inquire whether they could be related by a finite BRST transformation.  That final missing point is fulfilled with an affirmative answer in the next section.

\section{Gauge-Fixing Correspondence via FFBRST}
The remarkable BRST symmetry has been around since the early groundbreaking works of Becchi, Rouet and Stora \cite{Becchi:1974xu, Becchi:1974md}
and Tyutin \cite{Tyutin:1975qk},
furnishing an essential tool for the safe quantization of gauge invariant theories.  A recent comparative account on standard BRST and its many related offspring symmetries, in terms of both Lagrangian and Hamiltonian quantization viewpoints, can be seen in \cite{Mandal:2022zil}.  In principle, the original BRST parameter, as $\omega$ in equation (\ref{BRS}), is infinitesimal, anti-commuting and global in nature.
The insightful idea of generalizing it to a finite field-dependent mathematical function, leading to finite field-dependent BRST (FFBRST) transformations, first appeared in reference \cite{JM1} and, since then, has found many important applications from which we mention a sample of representative recent results \cite{Upadhyay:2010ww, DPM,URM,PM1,PM2,MPT}.

In this section, we show how the two previous gauge-fixed quantum actions $S_u = S_{SZ} + S_{ug} $ and $S_{tt}$ in equation (\ref{Stt}) can be connected in terms of a FFBRST transformation.  Note that a simple infinitesimal gauge transformation cannot do here for at least two reasons: (i) gauge transformations such as (\ref{gs}) are defined only for the classical fields and (ii) $S_u$ and $S_{tt}$ are not infinitesimally close to each other.  Hence, following the original ideas of Joglekar and Mandal \cite{JM1}, we start by generalizing (\ref{BRS}) to
\begin{equation}\label{FBRST}
\phi \longrightarrow \phi+(s\phi) \Theta'[\phi_\kappa]d\kappa
\end{equation}
where now $\Theta'[\phi_\kappa]d\kappa$ plays the role of a, still infinitesimal, anticommuting field-dependent quantity, expressed in terms of a continuous parameter $\kappa$.  The infinitesimal character of the transformation (\ref{FBRST}) is captured by the differential $d\kappa$, while the Grassmann function $ \Theta'[\phi_\kappa] $ assures the field dependence.  The finiteness of a genuine FFBRST transformation can then be achieved by integrating in $\kappa$ from $0$ to $1$ to produce
\begin{equation}\label{FFBRST}
\phi_{\kappa=1}=\phi_{\kappa=0}+(s\phi)\Theta[\phi]
\,,
\end{equation}
where now
\begin{equation}
\Theta[\phi]=\Theta'[\phi]\frac{\exp f[\phi] - 1}{f[\phi]}
\end{equation}
corresponds to the finite field-dependent parameter with $f[\phi]$ given by 
\begin{equation}
f[\phi] = \sum_\phi \int d^{26}x \frac{\delta \Theta'[\phi] }{\delta \phi } (s\phi)
\end{equation}
summing over all fields $\phi$.
As it happens, Joglekar and Mandal have shown in \cite{JM1} that a FFBRST transformation of the form (\ref{FFBRST}) leads to a nontrivial Jacobian in the integration functional measure
\begin{equation}
d\mu \rightarrow d\mu_\kappa = J(\kappa)d\mu
\,,
\end{equation}
which can be written as
a local functional of the fields in the generating functional exponential argument if and only if the condition
\begin{eqnarray}
\int d\mu_\kappa \left[\frac{1}{J(\kappa)}\frac{d J(\kappa)}{d \kappa}-i\frac{dS_1}{d\kappa}\right] e^{i \displaystyle (S_0 + S_1)} = 0 
\label{ffbc}
\end{eqnarray}
is satisfied. In (\ref{ffbc}), $S_0$ denotes a starting BRST-invariant action and $S_1$ a corresponding additional contribution in the path-integral resulting from the Jacobian of a FFBRST transformation.  Accordingly, $\frac{dS_1}{d\kappa}$ represents the total derivative of $S_1$ with respect to $\kappa$ in which the dependence on the fields $\phi_\kappa$ is also differentiated. The Jacobian change is then calculated as 
 \begin{equation}
 1 - \frac{1}{J(\kappa)}\frac{d J(\kappa)}{d \kappa} d k = \frac{J(\kappa)}{J(\kappa+d\kappa)} = \sum_{\phi}{\pm}\frac{\delta \phi_{\kappa+d\kappa}}{\delta \phi_\kappa}
\label{jacbc}
\end{equation}
where the symbol ${\pm}$ denotes a positive/negative sign for bosonic/fermionic fields ($\phi$), respectively.
In order to construct an appropriate  FFBRST  transformation establishing  the connection between the generating functionals associated to the
two previously discussed gauges for the Siegel-Zwiebach model,
we consider $S_0$ in (\ref{ffbc}) as given by $S_u = S_{SZ} + S_{ug} $ and choose the field-dependent function $\Theta^\prime$ in (\ref{FBRST})  as
\begin{eqnarray}
\Theta^\prime= i \gamma'\int d^{26} x \left[{\bar{ c}}_\mu \left(\frac{\xi b^\mu}{2}+\partial^\nu h^\mu_\nu - B^\mu\right) + \bar{ c}\left(\frac{\zeta b}{2m^2}+\frac{h}{2} - \theta \right)\right]
\label{thtp1} 
\,,
\end{eqnarray} 
with $\gamma'$ denoting an open adjusting constant and all the fields depending on $\kappa$.
Corresponding to this choice, the Jacobian variation is given by
\begin{eqnarray}
 \frac{1}{J(\kappa)}\frac{d J(\kappa)}{d \kappa} &=& - i \gamma'\int  d^{26} x \Big[ s {\bar{c}}_\mu  \left(\frac{\xi b^\mu}{2} + \partial^\nu h^\mu_\nu - B^\mu \right) + s \bar{c}\left(\frac{\zeta b}{2m^2}+\frac{h}{2} - \theta \right) \nonumber\\&&+\,\, {\bar{c}}_\mu\left(\partial^\nu s h^\mu_\nu - s B^\mu\right)  + \bar{c} \left(s h/2 
  - s \theta \right)\Big]
  \,.
 \label{cjcb}
\end{eqnarray}
Next, following the usual FFBRST technique steps \cite{Upadhyay:2010ww, DPM,URM,PM1,PM2,MPT}, we consider an ansatz for the local functional of fields $S_1$ consisting of all possible terms arising from a FFBRST transformation written as
\begin{eqnarray}
S_1[\phi_\kappa, \kappa] &=& \int d^{26} x \Big[\xi_1 s({\bar{c}}_\mu)\frac{\xi b^\mu}{2}+\xi_2 s({\bar{c}}_\mu) \partial^\nu h^\mu_\nu + \xi_3 s({\bar{c}}_\mu) B^\mu +\xi_4s(\bar{c}) \frac{\zeta b}{2m^2} + \xi_5 s(\bar{c})\frac{h}{2} \nonumber\\ &+& \xi_6 s(\bar{c}) \theta + \xi_7 {\bar{c}}_\mu  \partial^\nu s (h^\mu_\nu) +  \xi_8 {\bar{c}}_\mu s (B^\mu)  + \xi_9 \bar{c} s\frac{h}{2}  + \xi_{10} \bar{c} (s\theta)  \Big]
\label{ansso}
\,,
\end{eqnarray}
where $\xi_i=\xi_i(\kappa)$, $i=1,\dots, 10$, denote ten $\kappa$-dependent parameters subjected to the initial condition $\xi_i (0) = 0$.  
To satisfy the condition prescribed by equation (\ref{ffbc}), we calculate
\begin{eqnarray}
\frac{dS_1}{d\kappa} &=& \int d^{26} x \Big\{ \xi'_1 s({\bar{c}}_\mu)\frac{\xi b^\mu}{2}+\xi'_2({\bar{c}}_\mu) \partial^\nu h^\mu_\nu +  \xi'_3 s({\bar{c}}_\mu) B^\mu +\xi'_4s(\bar{c}) \frac{\zeta b}{2m^2}+ \xi'_5 s(\bar{c})\frac{h}{2} + \xi'_6 s(\bar{c}) \theta \nonumber\\ &+& \xi'_7 \partial^\nu s (h^\mu_\nu) {\bar{c}}_\mu\ +  \xi'_8 s (B^\mu){\bar{c}}_\mu  + \xi'_9 s\frac{h}{2} \bar{c} + \xi'_{10} (s\theta) \bar{c}  + \Theta'\Big[ \xi_2 s (\partial^\nu h^\mu_\nu) s({\bar{c}}_\mu)+ \xi_3 s({\bar{c}}_\mu) s(B^\mu) \nonumber\\&+& \xi_5 s(\frac{h}{2})s(\bar{c}) + \xi_6 s\theta s(\bar{c}) - \xi_7 s{\bar{c}}_\mu \partial^\nu s (h^\mu_\nu) - \xi_8 s{\bar{c}}_\mu s(B^\mu) + \xi_9 s\bar{c}s\frac{h}{2} + \xi_{10} s\bar{c} (s\theta)  \Big] \Big\}
\,,
\label{csowk}
\end{eqnarray}
where $\displaystyle\xi'_i = \frac {d\xi_i}{d\kappa}$.

Hence, substituting the results from equations (\ref{csowk}) and (\ref{cjcb}) into condition (\ref{ffbc}) we obtain the relation
\begin{eqnarray}
 &&\displaystyle\int  d\mu_\kappa \exp[i (S_0 + S_1[\phi_\kappa, \kappa])]\int d^{26} x \bigg\{ ( \gamma' + \xi'_1)s({\bar{c}}_\mu)\frac{\xi b^\mu}{2}+( \gamma' + \xi'_2)s({\bar{c}}_\mu) \partial^\nu h^\mu_\nu + (-\gamma' + \xi'_3)s({\bar{c}}_\mu) B^\mu \nonumber\\ &&+ (\gamma' + \xi'_4)s(\bar{\cal C}) \frac{\zeta b}{2m^2} + (\gamma' + \xi'_5)s(\bar{c})h +(-\gamma'+\xi'_6) s(\bar{c}) \theta +(\gamma' +\xi'_7)\partial^\nu s(h^\mu_\nu) {\bar{c}}_\mu + (-\gamma'+ \xi'_8)s(B^\mu){\bar{c}}_\mu  \nonumber\\ &&+ (\gamma' + \xi'_9)s (h) \bar{c} + (-\gamma'+ \xi'_{10})(s\theta) \bar{c} +\Theta' \Big[ (\xi_2 - \xi_7)\partial^\nu s(h^\mu_\nu) s({\bar{c}}_\mu) + (\xi_3 - \xi_8)s(B^\mu)s({\bar{c}}_\mu) \nonumber\\ &&+ (\xi_5 - \xi_9)s(\frac{h}{2})s(\bar{c}) + (\xi_6 - \xi_{10})s(\theta)s(\bar{c})  \Big] \bigg\} = 0
 \label{ffbrc}
 \,,
\end{eqnarray} 
from which it is clear that the nonlocal terms, proportional to $\Theta'$, vanish independently  if
\begin{equation}
\xi_2 - \xi_7 = \xi_3 - \xi_8 = \xi_5 - \xi_9 = \xi_6 - \xi_{10} = 0 \,,
\label{reiji}
\end{equation}
while the remaining local terms in (\ref{ffbrc}) vanish if the following conditions hold:
\begin{eqnarray}
\gamma' + \xi'_1 &=&0, \quad \gamma' + \xi'_2 =0, \quad -\gamma' + \xi'_3 = 0, \quad \gamma' + \xi'_4 = 0, \quad \gamma' + \xi'_5 = 0\,,\nonumber\\
-\gamma' + \xi'_6 &=&0, \quad \gamma' + \xi'_7 =0, \quad -\gamma' + \xi'_8 = 0, \quad \gamma' + \xi'_9 = 0, \quad -\gamma' + \xi'_{10} = 0\,.
\label{rbjg}
\end{eqnarray}
Solving the differential equations for $\xi_i(\kappa)$ with initial conditions $\xi_i(0) = 0$, we obtain the results
\begin{eqnarray}
\xi_1 = -\gamma' \kappa, \quad  \xi_2 = -\gamma' \kappa,\quad \xi_3 = \gamma' \kappa,\quad \xi_4 = -\gamma' \kappa,\quad \xi_5 = -\gamma' \kappa,\nonumber\\
\xi_6 = \gamma' \kappa, \quad \xi_7 = -\gamma' \kappa, \quad \xi_8 = \gamma' \kappa, \quad \xi_9 = -\gamma' \kappa, \quad \xi_{10} = \gamma' \kappa \,.
\label{voj}
\end{eqnarray}
which can then be inserted back into equation (\ref{ansso}).  Withal, by choosing the ansatz parameter $\gamma' = -1$, we write
\begin{eqnarray}
S_1[\phi_{\kappa=1}, \kappa=1]
&=& \int d^{26} x \Big[ s({\bar{ c}}_\mu)\frac{\xi b^\mu}{2} + s({\bar{c}}_\mu) \partial^\nu h^\mu_\nu - s({\bar{c}}_\mu) B^\mu + s(\bar{c}) \frac{\zeta b}{2m^2} + s(\bar{c})\frac{h}{2} \nonumber\\ &&- s(\bar{c}) \theta - \partial^\nu s (h^\mu_\nu) {\bar{c}}_\mu  + s (B^\mu){\bar{c}}_\mu - s (\frac{h}{2}) \bar{c}+ (s\theta) \bar{c} \,\, \Big]
\label{yeso}
\,,
\end{eqnarray}
from which, by using equations (\ref{BRST}), it is clear that 
\begin{equation}
S_{u} + S_1 =  S_{tt}
\,.
\end{equation}
Therefore, we have been able to prove that those two gauge-fixings are connected by a FFBRST transformation justifying the obtained graviton propagator (\ref{G}) for the small mass regime.

\section{Conclusion}
We have shown how the divergence problem of the massive symmetric rank two tensor field propagator in the  string critical dimension can be rigorously addressed within the FFBRST transformations framework.  Our approach has proved to lead to an important result for the small mass graviton propagator in linearized gravity which has been found to be finite in the massless limit in a genuinely attainable traceless transverse gauge.  Hence, we confirm the heuristic results advocated by Park and Lee in \cite{HPTL} and provide the corresponding necessary rigorous quantization steps in terms of the well-established BRST formalism.   By means of the FFBRST technique, we have connected a Fierz-Pauli type gauge to a corresponding class of new gauges on an equivalent model containing a vector and a scalar fields, arising from the 
open string action fourth mass level expansion.  The finite field-dependent BRST transformations have generated a nontrivial Jacobian in the quantum generating functional path-integral integration measure which has been responsible for the proper connection of the aforementioned gauges.  As a bonus, we have obtained a whole new class of gauges continuously depending on two open parameters.  The powerfulness of FFBRST transformations have been attested once more, as they have been able to connect the new class of gauge-fixings proposed here to the unitary one.
The present work suggests future directions regarding other similar analysis coming from different alternative string theory formulations and possible canonical analysis of higher-order terms of the SZ action.

\end{document}